\documentclass[draft,12pt]{article}
\usepackage{amsmath,amsfonts,palatino,amsthm,epsf}
\theoremstyle{definition}

\newcommand{\beq}{\begin{equation}}
\newcommand{\eeq}{\end{equation}}

\newcommand{\f}{\begin{equation}}
\newcommand{\ff}{\end{equation}}
\begin{document}
\title{The weak anthropic principle and\\ the landscape of string theory}
\author{George Ellis\thanks{Email address: george.ellis@uct.ac.za}\\
Mathematics Department,
University of Cape Town,\\
Rondebosch, Cape Town 7701, South Africa
\\and\\
Lee Smolin\thanks{Email address: lsmolin@perimeterinstitute.ca}\\
Perimeter Institute for Theoretical Physics,\\
Waterloo, ON N2J 2W9, Canada}
 \date{\today}
\maketitle
\begin{abstract}
We note that there is an exception to the general arguments that no
falsifiable predictions can be made, on the basis of of presently
available data, by applying the weak anthropic principle (WAP) to
the landscape of string theory. If there are infinitely more vacua
in the landscape for one sign of a parameter than the other, within
an anthropicaly allowed range, then under very weak assumptions
about the probability measure one gets a firm prediction favoring
that sign of that parameter. It is interesting to note that while
the understanding of the string landscape is evolving, present
evidence on the nature of the landscape allows such an argument to
be made, leading to the conclusion that the WAP favors a negative
value for the cosmological constant, $\Lambda$, in contradiction to
the result of astronomical observations.  The viability of applying
the WAP to string theory then requires that either there are found
an infinite discretum of anthropically allowed vacua for $\Lambda
>0$, or the recently found infinite discretum of solutions for
$\Lambda <0$ be reduced to a finite value.
\end{abstract}
\newpage

\section{Introduction}

Recently it has been argued by ourselves
\cite{lotc,ls-essay,ellsto,encyc} and others  that the present
dominant interpretation of the Weak Anthropic Principle (WAP),
whereby the existence of life is explained by random selection from
an ensemble of universes with differing properties
\cite{rees,weinberg}, is by itself not falsifiable, because any
predictions made from it depend on assumptions made about the
ensemble of universes and the probability measure taken on it. There
is a great deal of freedom in the choices that can be made here, and
those assumptions determine the expectations for experiments; but
they are untestable, in particular because all the other universes
in the supposed ensemble are unobservable \cite{ellsto,encyc}.

These considerations are the focus of present interest  because of
the claims that the WAP can be applied to the ``landscape"
\cite{lotc} of string theory \cite{Susskind,sus05}. According to
recent results, the landscape consists of an infinite set of
discrete vacua, which are argued to be different possible
quasistable ground states for string theory \cite{KKLT,DGKT,STW,AD}.
These include on the order of $N_{+}=10^{500}$
 flux compactifications  which appear compatible with 3+1 non-compact
dimensions and a positive cosmological constant $\Lambda$ \cite{KKLT}. This
landscape of possibilities leads to a vast number of different
predictions for post standard model physics \cite{KKLT}, rather than
a single prediction that can be verified or disproved. This has led to worries
that string theory fails to be predictive.

Combining these considerations, it appears at first that the WAP
places no constraints on string theory because it is expected that
the Landscape will include many versions of local physics that will
allow life to exist; consequently the WAP can always be fulfilled in
any ensemble of universes where most of the possibilities of the
Landscape are realised. The project of combining the WAP with the
landscape of string theory, as proposed in  \cite{Susskind,sus05},
then has been argued to be untestable.

In reply, it has been proposed that the theory makes exactly one
prediction,  which is that the sign of the spatial curvature be
negative \cite{sus05,fre05}. This in principle could allow disproof
of most versions of the theory, were the scalar curvature measured
to be positive. However there will for the foreseeable future be
considerable uncertainties that will prevent many from accepting
this kind of data as knock-out disproof of either the landscape or
the idea of an ensemble of universes, even if the observations come
out on the side of positively curved spatial sections.\footnote{ The
best present value of the spatial curvature parameter from the
combined astronomical data has  shifted from $\Omega_k =
1.02 \pm 0.02$ \cite{WMAP}, marginally indicating positively curved
spatial sections, to $\Omega_k = 1.003\pm 0.010$ \cite{teg},
consistent with positive or negative values.}

In this comment we point out there is another important exception to
the conclusion that the combination of the WAP with the string landscape is untestable.
In spite of the general difficulties,
we show that there are possible
distributions of vacua that allow an argument based on
the WAP to be made which leads to
a prediction which is stable under a large number of
choices for the probability measure on the ensemble of universes.

One case in which this can occur is if there are a finite number of
anthropically allowed vacua for one sign of $\Lambda$, but an
infinite discrete set of anthropically allowed vacua for the other
sign. For example, suppose that there is an infinite discretum of
anthropically allowed vacua for $\Lambda >0$ and only a finite
discretum for $\Lambda < 0$. We can then argue that under a large
set of possible choices for probability weights the theory will
predict $\Lambda >0$.   This is because there would then be two
discrete sets of universes where life may  be possible: one, ${\cal
S}_-$ (with $\Lambda < 0$), is finite, and the other, ${\cal S}_+$
(with $\Lambda > 0$) is infinite. If we follow the logic of the WAP
argument by choosing randomly from the infinite set ${\cal S} \equiv
{\cal S}_+ \bigcup {\cal S}_-$ which is the union of them (with
$\Lambda$ taking all values), there is vanishing probability that we
will end up in the finite subset ${\cal S}_-$ of negative
cosmological constant universes.  Thus, in this case, we would
predict that $\Lambda >0$.  This would be most gratifying as that is
what is observed \cite{supernovae,supernovae1}.

At present the study of the landscape is evolving, and it is too
soon to draw definitive conclusions of this kind.  But, for what it
may be worth, it can be pointed out that at present the situation is
the reverse of that just sketched.  There are at present claimed to
be a large, if finite discretum of vacua with values of the
cosmological constant that are positive but less than anthropic
bounds \cite{KKLT}. There is even a conjecture that the set of
anthropically allowed vacua consistent with $\Lambda >0$ is bounded
by a finite number \cite{AD}. Thus, at present, the evidence is that
the cardinality of the $\Lambda >0$, anthropically allowed
discretum, which we may call, $N_+$, is large, but  finite.

At the same time, the present evidence is that
the set of string vacua with negative cosmological
constant is actually countably infinite \cite{DGKT,STW}.
As  is shown in \cite{DGKT}, the number of
string vacua appears to diverge as $\Lambda \rightarrow 0$ {\it from below.}
In the vacua studied in \cite{DGKT} there is a condition that the compactification
volume also be large, so the number anthropically allowed will be a proper subset
and, perhaps even finite. However in \cite{STW} evidence is found for an infinite set of
additional negative $\Lambda$ compactifications, including some whose numbers diverge
within finite ranges of macroscopic parameters.  At present, our understanding
is that not enough is known about this set to know whether all but a finite set can
be excluded on anthropic grounds.  Thus the present situation (which may certainly change)
is that the best estimate for the number of
negative cosmological constant anthropically allowed vacua, $N_-$,  is
 \f
 N_- = \infty.
\ff

If further studies should confirm these estimates, then there would
be important consequences for the viability of the landscape
approach to string phenomenology. This is because the combination of
the results and conjectures in the papers \cite{DGKT,STW,AD} lead,
if correct, to the conclusion that the weak anthropic principle
implies that the cosmological constant will be negative, which is in
contradiction to the astronomically determined value
\cite{supernovae,supernovae1}. Consequently, if the present
situation turns out to be correct\cite{apology}, we will have to conclude that
either the string theory landscape does not describe nature, or the
Anthropic style of argument whereby the existence of life is
explained by random selection from an ensemble of universes cannot
be valid (perhaps because the hypothesized ensemble of universes
does not in fact exist). But this is then problematic for both
understandings, because on the one hand the landscape of string
theory is claimed to be \emph{the} natural setting for applying the
anthropic style of argument \cite{Susskind}, and on the other,
anthropic arguments seem to be the only way to get reasonably unique
physical predictions out of the landscape of string theory. If the
combination does not work, as we argue here is implied by the
current best understandings of the landscape, then both components
are on weak ground.

The purpose of this note is to make this argument more carefully.

\section{The weak anthropic principle}

Here is a standard version of the  weak anthropic principle
\cite{BarrowTipler}.  We  posit that there is an actual ensemble
$\cal E$ of universes $u_i$, or of expanding universe domains in a
larger universe, which is a discrete set; we live in one of the
expanding universe domains in the ensemble. They may, for example,
have been produced by eternal inflation \cite{linde}, or other
mechanisms. We may take the number of actual expanding universe
domains $N_U$ to be arbitrarily large, or countably infinite.  On
$\cal E$ there is a measure $f_i$ which is proportional to the
probability for the universe $u_i$ to contain intelligent life. If
it is impossible for life to exist in a universe\footnote{From here
on we use `universe' as shorthand for `universe or expanding
universe domain'.} $u_i$, then $f_i=0$. There is then a subensemble
${\cal L} \subset {\cal E}$ of universes where intelligent life is
possible.

In some versions, including those associated with the {\it
principle of mediocrity} \cite{PM}, $f_i$ is proportional to
something like ``the number of civilizations" likely to exist in
the universe $u_i$. In other versions, $f_i$  has a constant
non-zero value if intelligent life is possible in $u_i$.

We may be interested in the value of an observable, ${\cal O}$ whose
value in the universe $u_i$ is ${\cal O}_i$.  According to the weak
anthropic principle its expectation value is given by \f < {\cal
O}>= \frac{\sum_i^\prime  {\cal O}_i f_i}{\sum_i^\prime f_i}
\label{expectation} \ff where the sum indicated by $\sum_i^\prime$
means sum over all the universes in the ensemble ${\cal L}$ where
intelligent life is possible.  Alternatively, let ${\cal Q}$ be some
property that a universe may or may not have. Those universe which
have the property live in an ensemble ${\cal E}_{\cal Q}$. Let
$u'_i$ denote all $u_i$ such that $u_i \in {\cal E}_{\cal Q} \cap
{\cal L}$. Then the probability that $\cal Q$ is true in a typical
randomly chosen universe with intelligent life is \f P_{\cal Q} =
\frac{\sum_{u'_i  f_i }}{\sum_i^\prime f_i} \ff The reason why it is
sometimes argued that the weak anthropic principle implies directly
no predictions is that these observables depend crucially on the
choices made of the weights $f_i$. However, given information about
the $f_i$, some predictions are possible, as we shall now see.

\section{Anthropic constraints on the cosmological constant}

A famous argument of Weinberg tells us that there is a maximum value
of the cosmological constant $\Lambda$ compatible with the existence
of galaxies \cite{weinberg}. Since it is believed that stars form in
galaxies, and planets providing viable habitats for life circle
around stars, we take this value, which we call $\Lambda_+$ as the
limiting value necessary for intelligent life.

There is, so far as we know, no cosmological or astronomical reason
that life is not compatible with zero cosmological constant. There
can therefor be no astronomical reason life is not compatible with a
small negative cosmological constant, but there will be a lower
limit $\Lambda_-$ to the negative values of $\Lambda$ compatible
with intelligent life, because all universes with a negative
cosmological constant recollapse, and too negative a value for
$\Lambda$ will imply extremely short life times for the universe
before they recollapse, not allowing the extended times needed for
the evolution of intelligent life .

There seems no reason why life is not roughly as probable in a
member of the string theory landscape with slightly negative value
of $\Lambda$ than slightly positive. It is true that for the positive
$\Lambda$ members of the discretum, supersymmetry is always broken
at the level of the effective supergravity description, whereas for
the negative $\Lambda$ members sometimes it is broken at that level
and sometimes it is preserved. It is also possible that
supersymmetry breaking is necessary for life. But even if that were
the case, there is a general expectation that even in a world
governed by a string theory compactification with $N=1$
supersymmetry, supersymmetry can break at a much lower scale.
Indeed, a  common expectation has been that a phenomenologically
realistic  string theory would be compactified with $N=1$
supersymmetry, leading to a low energy phenomenology given by the
MSSM (minimal supersymmetric standard model), and that supersymmetry
would be broken spontaneously at the weak scale. There seems no
reason to believe that this is more or less likely for members of
the discretum with slightly positive $\Lambda$ than slightly
negative $\Lambda$.

The main relevent differences between the two cases, from the point
of view of the weak anthropic principle, is that according to
current understandings of the landscape of string theory, there is a
finite set of the slightly positive $\Lambda$ elements of the
discretum, while the slightly negative $\Lambda$ discretum is an
infinite set.

We may then reason in the following way. Consider an actually
existing ensemble of universes which includes our own universe,
which is selected from the ensemble because it allows life to exist.
Let us define the function $f(\Lambda)$ to be the anthropic weight
given to a universe in the ensemble with cosmological constant
$\Lambda$, averaging over all other parameters. That is, it is the
probability that intelligent life will occur in that universe. We
require that $f(\Lambda )$ be a smooth, normalizable function such
that if $\Lambda_i$ is the cosmological constant of the universe
$u_i$, \f f_i = f(\Lambda_i) w_i \ff Here $w_i$ is a weight that
accounts for everything in the properties of the universe that
affect the existence of intelligent life, except the cosmological
constant. Assuming the existence of this function dependent only on
$\Lambda$ is equivalent to assuming that the probability for life to
occur is a separable function, so that we can consider the
dependence on $\Lambda$ separately from the dependence on other
parameters\footnote{For a discussion of parametrisation of universes
in relation to the anthropic principle, see \cite{ellsto}.}. This
may not be true for all circumstances, but is probably true for
universes close to ours in the ensemble of possibilities, that is,
those with properties similar to that in which we live. In any case
this is the assumption made by Weinberg in his analysis
\cite{weinberg}, and by most papers that followed that one. We will
see what the result is on this basis, taken as a first approximation
that allows an analysis of the probabilities. Further analyses could
look into weakening this restriction; we doubt that will change the
result.

For universes within the ensemble $\cal L$ of universes with life,
we can assume there are minimum and maximum values $w_{min}$ and
$w_{max}$ such that \f \forall u_i \in {\cal L}, \, 0 <  w_{min}
\leq w_i \leq w_{max}. \label{wcondition} \ff That is, given all
other factors, the expectation for life does indeed depend crucially
on the cosmological constant; but that dependence is not infinitely
sensitive.

There are many details that go into the $w_i$ whose effects on the
chance for intelligent life are difficult to estimate with present
knowledge, such as whether supersymmetry breaking happens at the
Planck scale or lower, or whether indeed supersymmetry breaking is
necessary for life.  But these details are not crucial; all we need
is the condition (\ref{wcondition}).

We  may also impose very weak conditions on $f (\Lambda ) $ coming
from our knowledge of these anthropic conditions. We define the
ratio \f r=\lim_{\epsilon \rightarrow 0+}
\frac{\int_{\Lambda_-}^{\epsilon}  f(\Lambda)\, d\Lambda
}{\int_{\epsilon}^{\Lambda_+}  f(\Lambda)\,d\Lambda} \, .\ff As
discussed above, there seems no reason that a slightly positive
value of the cosmological constant should be infinitely more
friendly to life than a slightly negative value. We then propose it
is reasonable to assume
\begin{itemize}

\item{\bf A:} $f(\Lambda )$ is smooth and $\int_{\Lambda_-}^{\Lambda_+} d\Lambda
f(\Lambda ) =1$.

\item{\bf B:}{ $r$ is a non-zero, but finite number.}

\end{itemize}

The first condition  gives a normalization for $f(\Lambda )$ and
implies that $f(\Lambda )$ is bounded.  The second says that the
conditions for intelligent life do
  not favor one sign of the cosmological constant over another
  sign more than can be expressed by a finite, non-zero ratio.
  Together with the first condition it implies that $f(\Lambda )$
  is bounded separately for positive and negative $\Lambda$.

  Given that there
is very unlikely to be an anthropic reason to favor very small
positive values of $\Lambda$ over very small negative values, these
seem reasonable assumptions that will in fact necessarily be
fulfilled without implying any further restrictions on the physics
considered. But they are the specific assumptions we will need in
what follows, and we label them as such. In effect they explicate
the anthropic requirements that have to be fulfilled in relation to
the values of $\Lambda$. Further analyses could look at weakening
these conditions.

\section{A prediction from combining string theory with the weak anthropic principle}

Given the definitions made above, we can compute the {\it
probability that the  cosmological constant is strictly greater than
some positive value $\epsilon < \Lambda_+$ } on the basis of the
properties of the string theory landscape. Suppose that selection of
the universe in which we live takes place in an ensemble associated
with such a landscape. Let $i(\epsilon,\Lambda_+)$ be the set of
values $i$ such that $\epsilon \leq \Lambda_i \leq \Lambda_+$ and
$i(\Lambda_-,\Lambda_+)$ be the set of values $i$ such that
$\Lambda_- \leq \Lambda_i \leq \Lambda_+$. Then on the basis of the
present understandings of the landscape of string theory, summarized
above, we argue as follows: \f P_{\Lambda
>\epsilon } = \frac{\sum_{i(\epsilon,\Lambda_+)
}^\prime f_i }{\sum_{i(\Lambda_-,\Lambda_+)}^\prime f_i } \ff For
any finite value of $\epsilon$ much smaller than the anthropic
limit $\Lambda_+$ the numerator is a finite sum, albeit over a
large number of elements.  But the denominator is an infinite sum.
Given the conditions {\bf A} and {\bf B} we have assumed for $f$
it follows that each term in the sums in the numerator and
denominator is finite, positive and bounded.  Then,
 for any positive $\epsilon$ much smaller
than $\Lambda_+$ we have\footnote{Here is the demonstration \f
P_{\Lambda >\epsilon } < \frac{w_{max}}{w_{min}} \\
\frac{\sum_{i(\epsilon,\Lambda_+) }^\prime f (\Lambda_i )
}{\sum_{i(\Lambda_-,\Lambda_+)}^\prime f (\Lambda_i )  } =0 .\ff
This vanishes because the $f(\Lambda )$ are bounded, so the
expression is a ratio of a finite sum of positive bounded terms to
an infinite sum of positive bounded terms. }. \f \Lambda_+\gg
\epsilon > 0 \Rightarrow P_{\Lambda
>\epsilon } = 0 \ff Alternatively we can compute the expectation
value of $\Lambda$. From (\ref{expectation}) we have \f <
\Lambda>= \frac{\sum_i^\prime f_i \Lambda_i  }{\sum_i^\prime f_i}
\label{expectation2} \ff It follows from our assumptions plus the
results on the string landscape \cite{DGKT,STW} that \f <\Lambda>~
\leq~ 0 . \ff This is because there are infinitely more universes
in (\ref{expectation2}) with $\Lambda \leq 0$ than for strictly
positive values.

These predictions disagree with current astronomical observations,
which show that $\Lambda > 0$ at present
\cite{supernovae,supernovae1}, with the observations being
compatible with the `dark energy' causing the observed acceleration
being a cosmological constant \cite{pad}.\footnote{Even if the
observations eventually show `dark energy' varies with time and so
is not a cosmological constant, the Landscape argument predicts
there will indeed be a negative cosmological constant! This is not
what is observed.} Thus, one can say that given the assumptions {\bf
A} and {\bf B} together with the current results on the string
theory landscape, by using the WAP we reach a prediction that is
falsified by current observations. This is the result claimed in the
introduction.

We can also consider the probability $P_{compatible}$ that string
theory is in the finite set, conjectured by Acharya and Douglas
\cite{AD} to be compatible with all current observations. Given that
this is a finite subset of string vacua, and that there is an
infinite set of string vacua not compatible with current
observations, one reaches the conclusion by similar reasoning that
\f P_{compatible} = 0. \ff This reinforces the conclusion reached
above. Similarly, we can consider the possibility that there is a
continuous infinite set of zero cosmological constant,
supersymmetric vacua.\footnote{David Gross, Private communication.}.
This would also strengthen the conclusion.

\section{Conclusions}

We have found that, while it is difficult generally to extract
falsifiable predictions from a combination of the WAP and the string
landscape, there are exceptions in which the statistics of the
string vacua are weighed so heavily towards one range of parameters,
that there are predictions which are stable under a wide range of
choices of probability weights.  As we have seen here, the present
data on the landscape, as of this date, allows such an argument to
be made with regard to the sign of the cosmological constant, and
that prediction happens to be in disagreement with observation.

It may very well be that this situation changes when the landscape
is better explored. There are four basic possibilities, depending on
whether $N_+$ or $N_-$ are finite or infinite. If both are finite or
both are infinite,  it may be difficult to get a firm prediction
which does not depend on otherwise untestable assumptions about the
probability weights.  But if one is finite and the other infinite,
as is the case with our present understandings, we can expect a firm
prediction, one of which would be right, in that it agrees with the
data, and the other wrong - it would indeed be falsified. The latter
is what is in fact implied by the current status of understanding of
the landscape.

\section*{ACKNOWLEDGEMENTS}

We are grateful to Stephon Alexander, Michael Dine, Shamit Kachru,
Washington Taylor and Jeff Murugan for correspondence and
discussions on this issue.  We are also grateful to Paul Davies for invitations to a workshop
at Arizona State University where these issues were discussed.  Research at Perimeter Institute for
Theoretical Physics is supported in part by the Government of Canada
through NSERC and by the Province of Ontario through MEDT. The
Cosmology Group at the University of Cape Town  is supported by the
NRF (South Africa) and the University of Cape Town Research
Committee.

\end{document}